\documentclass[prb,twocolumn,showpacs,floatfix,amsfonts]{revtex4}
\usepackage{graphicx,graphics,color,epsfig}
\usepackage{bm}
\usepackage{amsmath}
\usepackage{amssymb}
\begin{document}



\title{Magnetic Quantum Phase Transitions in Kondo Lattices}

\author{Qimiao Si}
\affiliation{Department of Physics \& Astronomy, Rice University,
Houston, TX 77005--1892, USA }

\author{Jian-Xin Zhu}
\affiliation{Theoretical Division, Los Alamos National Laboratory,
Los Alamos, New Mexico 87545, USA}

\author{D. R. Grempel}
\affiliation{ CEA-Saclay, DSM/DRECAM/SPCSI, 91191 Gif-sur-Yvette,
France }

\begin{abstract}
The identification of magnetic quantum critical points in heavy
fermion metals has provided an ideal setting
for experimentally studying 
quantum criticality. Motivated by these experiments,
considerable theoretical efforts have recently been devoted to
reexamine the interplay between Kondo screening and magnetic
interactions in Kondo lattice systems. A local quantum critical
picture has emerged, in which magnetic interactions suppress Kondo
screening precisely at the magnetic quantum critical point (QCP).
The Fermi surface undergoes a large reconstruction across the QCP
and the coherence scale of the Kondo lattice vanishes at the QCP.
The dynamical spin susceptibility exhibits $\omega/T$ scaling and
non-trivial exponents describe the temperature and frequency
dependence of various physical quantities. These properties are to
be contrasted with the conventional spin-density-wave (SDW)
picture, in which the Kondo screening is not suppressed at the QCP
and the Fermi surface evolves smoothly across the phase
transition. In this article we discuss recent microscopic studies
of Kondo lattices within an extended dynamical mean field theory
(EDMFT). We summarize the earlier work based on an analytical
$\epsilon$-expansion renormalization group method, and expand on
the more recent numerical results. We also discuss the issues that
have been raised concerning the magnetic phase diagram. We show
that the zero-temperature magnetic transition is second order when
double counting of the RKKY interactions is avoided in EDMFT.

\end{abstract}

\maketitle

\newpage

\section{Introduction}

Heavy fermions started out as a fertile ground to study strongly
correlated Fermi liquids and superconductors.~\cite{StewartRMP84}
There was a great deal of amazement 
at seeing 
a Fermi liquid whose
quasiparticle mass is over a hundred times
the bare electron mass; hence the name of the field.
It was also surprising to find superconductors in an inherently magnetic
environment: the existence of local magnetic moments in these materials
is established through the observation of a Curie-Weiss susceptibility
at intermediate temperatures (of the order of 100 K), and magnetism is
supposed to be ``hostile'' to
superconductivity. It was qualitatively understood that the large
mass reflects
the Kondo screening of the magnetic moments, which is necessary to overcome
magnetism,~\cite{Doniach77,Varma02} and the proper theoretical
description of the Fermi liquid state~\cite{Bickers88}
was subsequently achieved. When high temperature
superconductors were discovered in 1986, the development of the
heavy fermion field was naturally interrupted -- at least partially.
The hiatus proved to be relatively short-lived. As it re-emerged,
however, the
field acquired a considerably different outlook, with the emphasis now
placed on
non-Fermi liquid behavior and
magnetic quantum phase transitions.
Since the late 1990s the field has become a focal
point~\cite{StewartRMP01}
for the general study of quantum
criticality. The interest in QCPs is by no means unique to heavy
fermions; it also arises in high temperature superconductors among
other materials.~\cite{Varma02,Sachdev-science} However, heavy
fermions are particularly advantageous in one important regard.
That is,
second-order quantum phase transitions are explicitly
observed in a growing list of this family of materials.

The QCP in heavy fermions typically separates an antiferromagnetic
metallic phase from a paramagnetic metallic phase. Near the
magnetic QCP, transport and thermodynamical properties develop anomalies.
The $T=0$ SDW
picture~\cite{Hertz76,Millis,Moriya,Lonzarich-review}
describes the QCP in terms of fluctuations of
the magnetic order parameter -- the paramagnons -- both in space
and in time.
This theory amounts to a $\phi^4$ theory -- 
with $\phi$ being
the paramagnon field -- in an effective dimensionality
of
$d_{\rm eff}=d+z$.
Here,
$z$ is the dynamic
exponent,
and 
is equal to $2$ in the antiferromagnetic case.
So, 
$d_{\rm eff}$ 
is above the upper critical dimension,
$4$,
of the $\phi^4$ theory,
for spatial dimensions $d \ge 2$.
The corresponding fixed point is Gaussian.
In this picture, the non-Fermi liquid properties of
the QCP reflect the singular scattering of electrons by the
paramagnons
and
are unrelated to the process of Kondo screening.

The most direct indication for the unusual nature of the heavy
fermion quantum criticality came from inelastic neutron scattering
experiments.~\cite{Schroeder00,Aronson95} The magnetic dynamics
shows a fractional exponent and  $\omega/T$ scaling over an extended
range of momentum space. These features deviate drastically from the
expectations of the SDW theory. Since the SDW critical fixed point
is Gaussian, spin damping would be controlled by some
(dangerously) irrelevant coupling, and neither fractional
exponent nor $\omega/T$ scaling would be expected.

One way to resolve this impasse invokes the breakdown of Kondo
screening at the magnetic quantum critical point. On the
paramagnetic side, local moments become entangled with the
conduction electrons and, in the process, 
are 
delocalized and
a part of the electron fluid. At the magnetic QCP, the magnetic
fluctuations 
turn
soft and act as a source of dissipation that
couples to the local moments; this coupling competes with the Kondo
interactions and destroys  the Kondo screening.
Going from the paramagnetic side to the QCP,
the electronic excitations
depart from those of a Fermi liquid and acquire
a non-Fermi liquid form. These non-Fermi liquid excitations
are part of the quantum critical spectrum. Unlike the paramagnons,
they are characterized by an interacting fixed point.
As the control parameter is tuned further, into the magnetically
ordered side, the system is again a Fermi liquid, but the Kondo
effect is completely destroyed.
An important corollary of this picture~\cite{Si-Nature01} is that
the Fermi surface has a sharp jump at the QCP.

To microscopically study the magnetic quantum phase transition
requires an approach that can handle not only the heavy fermion
 and magnetic states but also the dynamical competition
between the two on an equal footing. One
 suitable approach (if not the
only one so far) is the extended dynamical mean field theory
(EDMFT).~\cite{Smith1,Smith2,Chitra,note}

In this article, we discuss the EDMFT studies of the Kondo
lattice system. In addition to a brief summary of the
earlier analytical works,~\cite{Si-Nature01} we will pay
particular attention to two issues. The first one
deals with the magnetic dynamics
near the QCP. A fractional dynamical-spin-susceptibility exponent
accompanies the destruction of Kondo screening.~\cite{GrempelSi03}
The second one concerns the nature of the zero-temperature transition.
When the EDMFT is implemented such that there is no double-counting
of the static component of the RKKY interactions, the transition
turns out to be
second order.~\cite{ZhuGrempelSi03} We compare
these results with those of some recent related
works.~\cite{SunKotliar03,SunKotliar05,Pankov04}

\section{The Model, the Phases, and the Phase Transitions}

\subsection{Kondo Lattice Model}

We consider a Kondo lattice model,
\begin{eqnarray}
{\cal H} &=& {\cal H}_f + {\cal H}_c + {\cal H}_K .
\label{kondo-lattice}
\end{eqnarray}
Here, the $f$-electron component
\begin{eqnarray}
{\cal H}_f &=&
\frac{1}{2}
\sum_{ ij}
I_{ij}^a
~S_{i}^a
~ S_{j}^a ,
\label{H-f}
\end{eqnarray}
 describes the interactions between spin-$\frac{1}{2}$ local
moments and $a=x,y,z$ are the spin projections. We have taken
the valence fluctuations to be completely
frozen, which should be a good description of at least those heavy fermion
metals that have a heavy effective mass and that are undergoing a
magnetic quantum phase transition. Without loss of generality, we
have assumed that a unit cell contains one local moment ($S_{i}$),
whose spin-$\frac{1}{2}$ nature reflects the projection onto the
lowest Kramers doublet. $I_{ij}^a$ describes the RKKY exchange
interaction between the local moments. In the physical systems,
the RKKY interaction
is generated by the Kondo interactions. Here, we have taken it as an
independent parameter, for two reasons. First, it is useful to do so
for the purpose of specifying the global phase diagram. Second, as
it will be seen
below, in the EDMFT approach to Kondo lattice,
it is necessary to include this parameter at
the Hamiltonian level to treat its effects dynamically (see
sections \ref{paramagnetic} and \ref{ordered} for details).

The conduction electron component of Eq.~(\ref{kondo-lattice})
is simply
\begin{eqnarray}
{\cal H}_c &=& \sum_{\bf k \sigma} \epsilon_{\bf k}
c_{{\bf k}\sigma}^{\dagger} c_{{\bf k}\sigma} .
\label{H-c}
\end{eqnarray}
It is implicitly assumed that the conduction electrons alone would
form a Fermi liquid, and the
residual interaction (Landau) parameters for the
conduction-electron component alone can be neglected. We will
consider the number of conduction electrons per unit cell, $x$, to
be in the range $0<x<1$; in addition, we will assume that it
is not too small, so that
the physical RKKY interaction between the nearest-neighboring local
moments is antiferromagnetic, and not too close to $1$, so that the
Kondo insulating physics does not come into play. All the phases
described below are metallic.

Finally, the local moments interact with ${\bf s}_{c,i}$,
the spins of the conduction electrons,
through an antiferromagnetic Kondo coupling $J_K$:
\begin{eqnarray}
{\cal H}_K &=& \sum_i J_K ~{\bf S}_{i} \cdot {\bf s}_{c,i} .
\label{H-K}
\end{eqnarray}

\subsection{Kondo Effect and Magnetic Quantum Phase Transition}

For qualitative considerations, consider the Kondo lattice model
with a fixed value of $I$ and $W$, with
a small ratio $I/W$; 
here $I$ is the
typical (say, nearest-neighbor) interaction of $H_f$, and $W$ the
conduction-electron bandwidth. We further assume that $H_{f}$ has an
Ising anisotropy. In the antiferromagnetic phase of $H_f$,
the spin excitation spectrum is fully gapped. An infinitesimal
$J_K$ cannot lead to any Kondo screening. Hence, the Fermi
surface encloses
only the conduction electrons, whose number is $x$
per paramagnetic unit cell. We label this phase AF$_{\rm S}$.

On the other hand, when $J_K$ dominates over $I$, the standard Kondo
screening does occur. Each local moment is converted into a
spin-$\frac{1}{2}$ charge $e$ Kondo resonance. The Fermi 
surface
now
encloses 
not only the conduction electrons but also the local moments,
the total number being $1+x$ per unit cell. We label this phase
PM$_{\rm L}$. While the existence of this phase is
well-established,~\cite{Bickers88} the easiest physical way to see
it is to consider the limit $J_K \gg W \gg I$. [Since the Kondo
state restores $SU(2)$ symmetry, we have, without loss of
generality, taken the Kondo exchange coupling to be spin-isotropic
in Eq.~(\ref{H-K}).] In this limit, there is a large binding energy
(of the order $-J_K$) for a local singlet between ${\bf S}_i$ and
${\bf s}_{c,i}$, and we can safely project to this singlet subspace.
In this subspace, $x=1$ becomes special: here each local moment is
locally paired up with a conduction electron, and the entire system
becomes a Kondo insulator. For a system of total $\mathcal{N}$ unit
cells, an $x<1$ amounts to creating $(1-x)\mathcal{N}$ unpaired
local moments, each of which is equivalent to creating a hole in the
singlet background. The Kondo lattice model becomes equivalent to an
effective single band Hubbard model of $(1-x)$ holes per site, with
an infinite on-site repulsion (it is impossible to create two holes
-- there is only one electron in the singlet to begin
with).~\cite{Lacroix85,Nozieres98}
In the paramagnetic phase, the
Luttinger theorem then ensures that the Fermi volume contains
$(1-x)$ holes or, equivalently, $(1+x)$ electrons!.

These general arguments show that the AF$_{\rm S}$ and PM$_{\rm L}$
phases are two stable metallic phases. They differ in two important
regards. The AF$_{\rm S}$ is magnetically ordered while the PM$_{\rm
L}$ is not. Equally important, the PM$_{\rm L}$ has the Kondo
screening while the AF$_{\rm S}$ does not. Increasing the ratio
$\delta \equiv J_K/I$ takes the system from AF$_{\rm S}$ to PM$_{\rm
L}$. A key question is this: Does the destruction of magnetism and
the onset of Kondo screening occur at the same stage? If so, the
transition is distinctly different from the $T=0$ SDW picture. If
not -- {\it i.e.}, if the destruction of magnetism happens after the
Kondo screening has already set in -- then the magnetic transition
can be interpreted as an SDW instability of the quasiparticles near
the large Fermi surface; the transition goes back to the realm of
the $T=0$ SDW transition picture.

Microscopical studies provide a way to address this issue. A
suitable method has to capture not only the AF$_{\rm S}$ and
PM$_{\rm L}$ phases, due respectively to the RKKY exchange interactions
and the Kondo interactions, but also the dynamical competition between
these interactions; this dynamical interplay is crucial for the
transition region. At this stage, the EDMFT method is the only one
we are aware of which fits this requirement.

Of course, microscopic studies always have limitations.
Approximations are inevitably used, in the process of solving
a Hamiltonian or at the level of the model itself (or both).
Controlled approximations, nonetheless, provide us 
with not only
ways
of understanding experiments but also intuitions that
serve as a basis for more macroscopic approaches. This general
philosophy is readily
reflected in the EDMFT approach. Even though it is
``conserving'' ({\it i.e.}, satisfying thermodynamic consistency
requirements),
it assumes that the 
${\bf q}$-dependences 
of some irreducible
single-electron and collective quantities [$M(\omega)$
and $\Sigma (\omega)$ defined in the next section] are
unimportant.

There are several reasons to believe that the EDMFT approach is
useful for the antiferromagnetic quantum transitions at hand. First,
because staggered magnetization is not a conserved quantity, the
spin damping does not have to acquire a strong dependence on ${\bf
q}$. This is especially true in metallic systems, where the dynamic
exponent $z$ associated with the long-wavelength magnetic
fluctuations is larger than 1. In the SDW case,~\cite{Hertz76} for
instance, $z=2$ and $M({\bf q}, \omega)$ can simply be taken as a
linear function of $|\omega|$ without any singular dependence on
${\bf q}$. Whether the ${\bf q}$-dependence in $M({\bf q}, \omega)$
is singular or regular can be stated in terms of the anomalous
spatial dimension $\eta$ characterizing the long-wavelength
fluctuations in space: a non-zero $\eta$ means that the ${\bf q}$
dependence in $M({\bf q}, \omega)$ is more singular than that
[$({\bf q}-{\bf Q})^2$] already incorporated in $I_{\bf q}$. So, if
$\eta \ne 0$, the EDMFT is expected to fail, at least for the
asymptotic behavior. For instance, the classical critical points
associated with a finite temperature magnetic transition in $d=2,3$
must have a finite $\eta$; the EDMFT turns out to produce (an
artificial) first-order phase transition. For a quantum critical
point, on the other hand, long-wavelength fluctuations occur in
$d_{\rm eff}=d+z$ dimensions. There is then a greater likelihood for
the vanishing of the spatial anomalous dimension, in which case the
${\bf q}$-dependence of $M({\bf q}, \omega)$ is not singular and
neglecting it will not change the universal behavior.

Second, as we have already discussed, the different classes of
magnetic quantum critical points of a Kondo lattice can be
classified in terms of whether the Fermi 
surface 
(in the paramagnetic
zone), which is large in the paramagnetic metallic phase, stays
large as the QCP is crossed or becomes small by ejecting the local
moments. Such large {\it vs.} small Fermi 
surfaces
are well described
in terms of whether the Kondo effect is preserved or destroyed,
which, in turn, are readily captured by the EDMFT approach.

We now turn to the EDMFT 
studies of 
the Kondo lattice
model.

\section{Destruction of the Kondo Effect within EDMFT}
\label{paramagnetic}

\subsection{The EDMFT equations for the paramagnetic phase}
\label{sec:edmft-eq-para}

The EDMFT approach treats certain intersite (coherent and
incoherent) collective effects on an equal footing with the
local interaction effects. The EDMFT equations have been
constructed
in terms of a ``cavity'' method,~\cite{Smith1} diagrammatics,~\cite{Smith2}
and a functional formalism.~\cite{Chitra}
All of these constructions yield the same dynamical equations.
In the diagrammatic language, the EDMFT is
seen as entirely different from a systematic
expansion~\cite{Schiller95,Georgesetal96} in $1/d$
whose zero-th order would correspond to the dynamical mean field
theory (DMFT).~\cite{Georgesetal96,MetznerVollhardt} Instead, the EDMFT is a
re-summation scheme that incorporates an infinite series of
processes associated with the intersite collective effects,
in addition to
the local processes already taken into account in the DMFT.
Unlike the single-electron properties, the collective modes do not
have a chemical potential. In other words, the bottom of a ``band''
is important and this provides a means for spatial dimensionality to
come into play in the EDMFT.

Within the EDMFT, the collective effects are organized in terms of
an explicit intersite interaction term at the Hamiltonian level. For
the Kondo lattice model described in the previous section, this is
the intersite exchange term,
${\cal H}_f$ of Eq.~(\ref{H-f}).

There are several ways to see the details of this formalism. One way
is to focus on a spin cumulant,~\cite{Smith2} whose inverse,
$M(\omega)$, is colloquially referred to as a spin self-energy
 matrix.
While it can be rigorously defined for any spinful many-body
problem, this quantity is taken as ${\bf q}$-independent in the
EDMFT.

The dynamical spin susceptibility, on the other hand, is
${\bf q}$-dependent and is given by
\begin{eqnarray}
\chi^a ({\bf q}, \omega) &=& \frac {1}  { M^a(\omega) + I^a_{{\bf q}}} .
\label{chi-edmft}
\end{eqnarray}
The conduction
electron self-energy is still given by $\Sigma (\omega)$, and the
conduction electron Green's function retains the standard form,
\begin{eqnarray}
G ({\bf k}, \epsilon) &=& \frac {1}  {\epsilon + \mu
- \epsilon_{\bf k} - \Sigma (\epsilon)} .
\label{g-edmft}
\end{eqnarray}
The irreducible quantities, $M(\omega)$ and $\Sigma (\epsilon)$, are
determined in terms of a self-consistent Bose-Fermi Kondo
model:
\begin{eqnarray}
{\cal H}_{\text{imp}} &=& J_K ~{\bf S} \cdot {\bf s}_c +
\sum_{p,\sigma} E_{p}~c_{p\sigma}^{\dagger}~ c_{p\sigma}
\nonumber \\
&& + \; g \sum_{p,a} S^a ~\left( {\phi}_{p,a} +
{\phi}_{-p,a}^{\;\dagger} \right) + 
\sum_{p,a}
w_{p,a}\,{\phi}_{p,a}^{\;\dagger}~ {\phi}_{p,a}\;. \nonumber
\\ \label{H-imp}
\end{eqnarray}
The self-consistency reflects the translational invariance:
\begin{eqnarray}
\chi_{{loc}}^a (\omega) &=& \sum_{\bf q} \chi^{a} ( {\bf q},
\omega ) ,
\nonumber \\[-1ex]
\label{self-consistent} \\[-1ex]
G_{{loc}} (\omega) &=& \sum_{\bf k} G( {\bf k}, \omega )\;.
\nonumber
\end{eqnarray}
When combined with the Dyson equations,
$M^{a}(\omega)=\chi_{0,a}^{-1}(\omega) + 1/\chi_{\rm loc}^{a}(\omega)$
and $\Sigma(\omega)=G_0^{-1}(\omega) - 1/G_{\rm loc}(\omega)$, where
$\chi_{0,a}^{-1} (\omega) = -g^2 \sum_p 2 w_{p,a} /[\omega^2 -
w_{p,a}^2]$ and $G_0 (\omega) = \sum_p 1/(\omega - E_p)$, are the
Weiss fields, these self-consistency equations specify the
dispersions, $E_{p}$ and $w_{p,a}$, and the coupling constant $g$.

We will focus on the case of two-dimensional magnetic fluctuations,
characterized by the RKKY density of states
$\rho_{I} (\epsilon) \equiv  \sum_{\bf q} \delta ( \epsilon  - I_{\bf q} )
= (1/{2 I}) \Theta(I - | \epsilon | )$.
The first of Eq.~(\ref{self-consistent}) becomes
\begin{eqnarray}
M^a(\omega) &=& I / {\rm tanh}[I \chi_{\rm loc}^a(\omega)] \nonumber \\
&=& I + 2I \exp \left[ -2I {\chi_{\rm loc}^a(\omega)} \right] + \cdots \;.
\label{M-expansion}
\end{eqnarray}
where the last equality is an expansion in terms of
$\exp
\left[ -2I  {\chi_{\rm loc}^a(\omega)} \right]$, valid
when the local susceptibility is divergent.

\subsection{Destruction of the Kondo effect}

Both the Kondo screening and its destruction are encoded in the
Bose-Fermi Kondo model, Eq.~(\ref{H-imp}). The antiferromagnetic
Kondo coupling 
($J_K$)
is responsible for the formation of a Kondo singlet
in the ground state and the concomitant generation of a Kondo
resonance in the excitation spectrum. The coupling of the local
moment to the dissipative bosonic bath 
($g$)
provides a competing
mechanism. To see this in some detail, we first analyze
Eq.~(\ref{H-imp}) alone without worrying about the
self-consistency conditions. We consider a given spectrum of the
bosonic bath,
\begin{eqnarray}
\sum_{p} [\delta(\omega - w_{p,a}) - \delta(\omega + w_{p,a})]
\propto |\omega|^{1-\epsilon} {\rm sgn} \omega \;.
\label{bosonic-bath}
\end{eqnarray}
The problem can be studied using an
$\epsilon$-expansion.~\cite{ZhuSi02}
For small $g$, the Kondo coupling 
dominates, leading to a Kondo screening. A sufficiently large
coupling $g$ destroys the Kondo screening completely, reaching a
local-moment phase. The transition between these two phases is of
second-order, and is described by a QCP where the Kondo screening is
just destroyed and the electronic excitations have a non-Fermi
liquid form.
The local spin susceptibility has a Pauli form on the Kondo side.
The destruction of Kondo screening is then manifested in a divergent
local susceptibility at the QCP. An important property that is shared
by the Bose-Fermi Kondo model with $SU(2)$ spin symmetry, $XY$ spin
anisotropy, or Ising spin anisotropy, is that
$\chi_{{\rm loc}}^a (\tau) \sim 1/\tau^{\epsilon}$,
at the QCP.
Here, $a=x,y,z$, $a=x,y$, and $a=z$ for the $SU(2)$, $XY$, and Ising
cases, respectively.

Correspondingly,
\begin{eqnarray}
\chi_{\rm loc}^a (\omega) = A^a(\epsilon)/(-i\omega)^{1-\epsilon} .
\label{chi-qcp}
\end{eqnarray}
While the critical amplitude,
$A^a(\epsilon)$,
depends on the spin anisotropy, the critical exponent does not;
it is equal to $1-\epsilon$ in all cases.

There is an important point that follows from the above analysis
which we will
use in the following discussion of the numerical results. Within the
EDMFT approach to the Kondo lattice, if the local spin
susceptibility of the Kondo lattice model is divergent at the
magnetic QCP, the corresponding local problem is sitting on the
critical manifold. In other words, a divergent local susceptibility
is a signature of the critical Kondo screening and its associated
non-Fermi liquid electronic excitations.

These $\epsilon$-expansion results for the Bose-Fermi Kondo model
were initially used to study the full self-consistent
problem,~\cite{Si-Nature01} with the self-consistency conditions
specified in the previous section. The fact that the critical
exponent for the local susceptibility is equal to $1-\epsilon$
[Eq.~(\ref{chi-qcp})] turns out to be essential for the existence of
a local QCP solution. The self-consistent solution in the case of
two-dimensional magnetic fluctuations has $\epsilon=1^-$,
corresponding [Eq.~(\ref{chi-qcp})] to a logarithmically divergent
local susceptibility.

\begin{figure}[t]
\centerline{\psfig{file=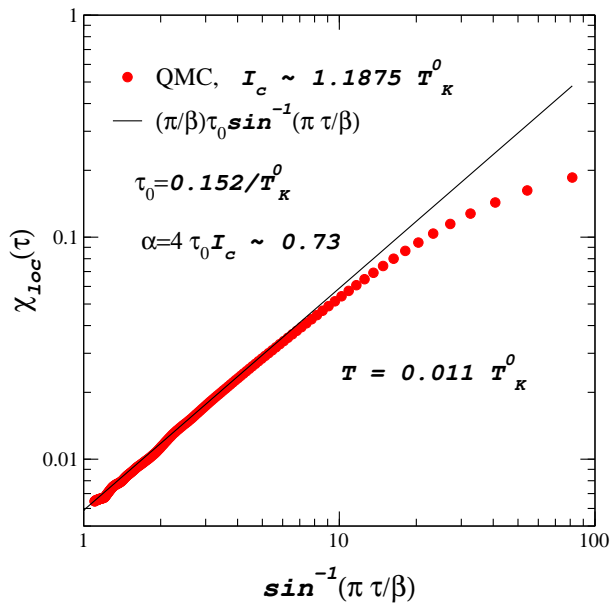, width=8cm}}
 \caption{The local spin susceptibility $\chi_{\rm loc}$ at the
quantum critical coupling ($I \approx I_c$)
as a function of the imaginary time $\tau$.
The log of $\chi_{\rm loc}$
is plotted against the log of $\sin^{-1}(\pi \tau /\beta)$.
The long-time limit, $\tau \rightarrow \beta/2$, corresponds to
$\sin^{-1}(\pi \tau /\beta) \rightarrow 1$. The solid line is a fit
in terms of $(\pi/\beta)\tau_0 \sin^{-1}(\pi \tau /\beta)$.
The fit yields a dynamical spin susceptibility exponent
$\alpha \approx 0.73$.
} \label{chi-tau}
\end{figure}

\begin{figure}[t]
\centerline{\psfig{file=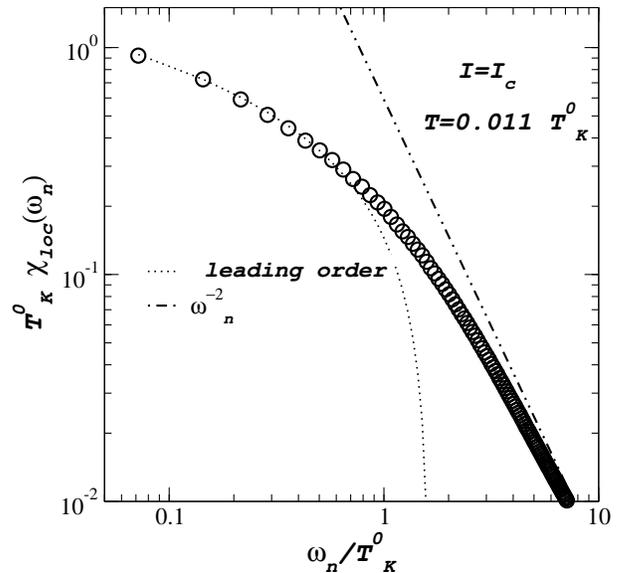, width=8cm}}
 \caption{Plot (log-log) of
the local spin susceptibility, $\chi_{\rm loc}$ {\it vs.} the
Matsubara frequency, $\omega_n$, at both low frequencies and high
frequencies. The dotted curve marked ``leading order'' corresponds
to a logarithmic dependence of $\chi_{\rm loc}$ on frequency. The
dot-dashed curve describes the fitting at high frequencies (not
shown); the $\omega_n^{-2}$ dependence is dictated by the spectral
sum rule. } \label{chi-omega}
\end{figure}

\subsection{Fractional exponent}

The EDMFT equations in the Ising case (taking only the $a=z$
component) were studied numerically in
Ref.~\onlinecite{GrempelSi03,ZhuGrempelSi03}
using 
the Quantum
Monte Carlo method of Grempel and
Rozenberg.~\cite{GrempelRozenberg99,Grempel98}

It was found that, at the magnetic QCP, the local spin susceptibility
is indeed logarithmically divergent.
Figure~\ref{chi-tau} shows a log-log plot of the
 local spin susceptibility $\chi_{\rm loc}(\tau)$ {\it vs.}
$\sin\left(\pi\tau/\beta\right)$ at a relatively low temperature
($T=0.011T_K^0$). It is seen from the figure that the zero-temperature limit
of the local susceptibility is $\chi_{\rm {loc}} (\tau)= A/\tau$. This
corresponds to a frequency dependence that is logarithmically
divergent in the low-frequency limit. A fit of the data yields
the value of the
 amplitude $A$ that is directly related to the critical exponent of
the peak value of the lattice susceptibility.

Figure~\ref{chi-omega} shows
the logarithmic dependence of $\chi_{\rm loc}(\omega_n)$ directly,
for frequencies smaller than the bare Kondo scale
$T_K^0$. As already mentioned in the previous
subsection, such a divergent local susceptibility signifies
that the Bose-Fermi Kondo model is located on the critical manifold;
correspondingly, there is a destruction of Kondo screening at
the magnetic QCP of the Kondo lattice model.

\begin{figure}[t]
\centerline{\psfig{file=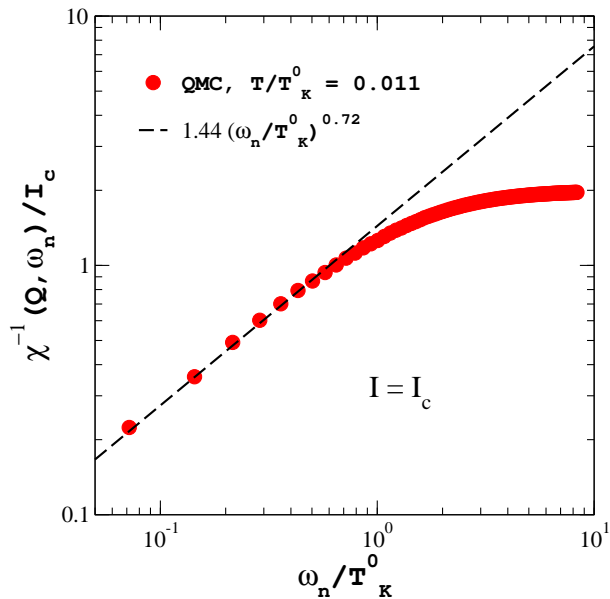, width=8cm}}
 \caption{Inverse peak susceptibility {\it vs}. Matsubara frequency,
at a low temperature ($T=0.011T_K^0$). We have used the
asymptotic form appearing in the second equality of Eq.~(\ref{M-expansion}).
The dashed line is a power-law fitting with the
exponent 0.72.} \label{peak-suscep}
\end{figure}

From the divergent local susceptibility, the self-consistency
equation~(\ref{M-expansion}) also determines the spin self-energy
and, by extension, the dynamical spin susceptibility of the Kondo
lattice. The inverse peak susceptibility, $\chi^{-1}({\bf
Q},\omega_n)$, where   ${\bf Q}$ is the ordering wavevector, is
shown in Fig.~\ref{peak-suscep}. A power-law fit yields a dynamical
spin susceptibility exponent that is fractional, close to $0.72$.

The fractional critical exponent is only seen at $|\omega_n| < T_K^0$.
Likewise, the Fermi-liquid
(linear in $\omega_n$) damping
inside the paramagnetic phase is also seen only at frequencies up to
at most $T_K^0$.

It is instructive to compare the above results with
those of Ref.~\onlinecite{SunKotliar03}, which studied
an Anderson lattice model. The lowest temperature studied in
Ref.~\onlinecite{SunKotliar03} is $0.25T_K^0$. The first
non-zero Matsubara frequency, $\omega_1=2 \pi T$,
is already larger than $T_K^0$. As a result, neither
the fractional exponent at the critical coupling $I \sim I_c$ nor the
linear damping at $I < I_c$ can be observed.

\subsection{Failure of the local $\phi^4$ description of the
Bose-Fermi Kondo model}

It is tempting to consider the Bose-Fermi Kondo model as equivalent to a
local $\phi^4$ theory. One maps the Kondo coupling to an Ising chain
(along the imaginary time $\tau $ axis) with $1/\tau^2$
interactions.~\cite{Andersonyuval} In addition, the bosonic bath,
with a spectrum of Eq.~(\ref{bosonic-bath}), adds an additional
retarded interaction, $1/\tau^{2-\epsilon}$. The corresponding
local $\phi^4$ theory is
$Z \sim \int {\cal D} \phi {\rm exp} [ -S]$ where
\begin{eqnarray}
S = \sum_{\omega_n} \left( r+ \frac{1}{g_c} |\omega_n|^2
+ \kappa_b |\omega_n|^{1-\epsilon}
+ \kappa_c |\omega_n|\right)\;\left|\phi(\omega_n)\right|^2
\nonumber\\
\label{local-phi4}
\end{eqnarray}
with the constraint $|\phi|^2=1$.
Indeed, such a local $\phi^4$ theory emerges in the large-$N$ limit
of a certain $O(N)$ generalization of the Bose-Fermi Kondo model.
In the $N=\infty$ case,
two of us~\cite{GrempelSi03} showed that, while the destruction
of Kondo screening does occur, the fractional exponent is absent.
Subsequently, Pankov {\it et al.}~\cite{Pankov04} demonstrated
that this conclusion remains valid to order $1/N$.

Recent works have considerably clarified the limitations of the large $N$
limit and demonstrated the failure of the local $\phi^4$ description
of the Bose-Fermi Kondo model, at least for $\epsilon \ge 1/2$ (including
the case of the self-consistent $\epsilon=1^-$). For the local-$\phi^4$
theory, $\epsilon=1/2$ is
the ``upper critical dimension''.~\cite{Fisher} At $\epsilon>1/2$,
the critical point would then be Gaussian, implying violations of
both $\omega/T$ scaling and hyperscaling. A number of recent studies on
the Bose-Fermi Kondo and closely related impurity
models~\cite{ZhuKirchner04,Vojta04,GlossopIngersent,Kirchner05}, have
shown that the quantum critical point is interacting over the entire
range $0<\epsilon<1$, obeying $\omega/T$ scaling and hyperscaling.
These results support the observation~\cite{GrempelSi03} of
$\omega/T$ scaling in the (self-consistent) case of
$\epsilon=1^-$. They also imply that the Bose-Fermi Kondo model is
perhaps the simplest model in which the standard description of a
QCP  -- in terms of a classical critical point in elevated
dimensions -- fails. Physically, the Kondo effect, involving the
formation of a Kondo singlet, is intrinsically quantum-mechanical.
In the language of a path integral representation for spin, the
Kondo singlet formation necessarily involves the Berry phase term.
It is then natural for the destruction of Kondo screening to be
inherently quantum mechanical and, by extension, for the QCP of
the Bose-Fermi Kondo model to be different from its classical
counterpart at a higher dimension. A more in-depth understanding
of the underlying mechanism for this effect will surely be 
illuminating.

\section{Magnetic Phase Diagram of Kondo Lattices within EDMFT}
\label{ordered}

We now address whether the above results, derived from the paramagnetic
side, are pre-empted by magnetic ordering. To do so, we approach the
transition from the ordered side.

\subsection{The EDMFT approach from the ordered side}

The EDMFT equations for the antiferromagnetically ordered phase
require normal ordering~\cite{Smith1,Smith2,Chitra} of $H_f$: $H_f =
\sum_{ ij} I_{ij} \left ( \frac{1}{2} ~:S_{i}^z: : S_{j}^z: +\langle
S_j^z\rangle S_i - \frac{1}{2} \langle S_i^z\rangle \langle S_j^z
\rangle \right )$, where the normal-ordered operator
is $:S_{i}^z: \equiv S_{i}^z -\langle S_{i}^z
\rangle$. The effective impurity model and the self-consistency
conditions are similar to Eqs.~(\ref{H-imp})
and~(\ref{self-consistent}), except for the following modifications.
First, there is a local magnetic field -- the static Weiss field,
$h_{\rm loc}$ -- coupled to $S^z$. This local
field,
arising through $I_{\bf Q}$, must be self-consistently
determined by the magnetic order parameter
 $\langle S^z\rangle_{\rm H_{imp}}$.

Second, the conduction electron propagators are also influenced
by magnetism.
It turns out that the second feature has to be treated with care so
that there is no double-counting of the RKKY interactions between
the local moments.
In Ref.~\onlinecite{ZhuGrempelSi03}, we avoided double-counting the
RKKY interactions by working with a featureless conduction electron
band; in this case, the magnetism is driven by the interaction
$I_{\bf q}$ already incorporated at the Hamiltonian level (in
$H_{f}$).
We will expand on this issue in the next Section.

\begin{figure}[t]
\centerline{\psfig{figure=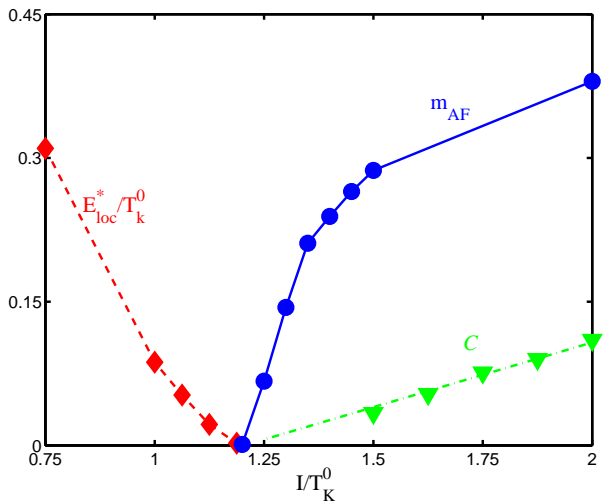,width=8.0cm,angle=0}}
\caption{ The coherence scale of the paramagnetic heavy-fermion
phase and the magnetic order parameter, $m_{\rm AF}$, {\it vs.} the
tuning parameter, $\delta \equiv I/T_K^0$. Both quantities are
determined at $T=0.011T_K^0$. Also shown is the Curie constant of
an unphysical solution without a magnetic order parameter at
$\delta
> \delta_c$. Lines are guides to eye. The fact that all three
curves meet at $\delta_c$ implies that the zero-temperature
transition is continuous. } \label{phase-diagram}
\end{figure}

Our phase diagram is shown in Fig.~\ref{phase-diagram}.
At $I<I_c$, the system is in the paramagnetic metal phase. The
coherence scale of the Kondo lattice $E_{\rm loc}^\star$ marks the
temperature/energy below which Kondo resonances are generated and
the heavy Fermi liquid behavior occurs. In particular, the Landau
damping is linear in frequency at $|\omega_n| < E_{\rm loc}^\star$. At
$I>I_c$, the system has an antiferromagnetic ground state. There is
a finite-temperature first-order transition at the N\'eel
temperature, $T_N$. However, $T_N$ continuously goes to zero as the
RKKY interaction is reduced. Within the numerical accuracy,
it vanishes as $I \rightarrow I_c^+$,~\cite{ZhuGrempelSi03} the same
place where $E_{\rm loc}^\star$ does so 
as $I \rightarrow I_c^-$. Plotted in
Fig.~\ref{phase-diagram} is the magnetic order parameter, $m_{\rm
AF}$, at the lowest studied temperature, $T=0.011T_K^0$. Again, it
is seen to continuously go to zero as $I \rightarrow I_c^+$.

Further support for the second order nature comes from the study of a
nominally paramagnetic solution at $I>I_c$. This solution
to the paramagnetic EDMFT equations co-exists with the solution to the
ordered EDMFT equations at $I>I_c$. But this ``paramagnetic'' solution
is found to contain
a Curie component $C/T$ in the static local susceptibility with
$C=\lim_{T\rightarrow 0} T \chi_{\rm loc}(T,\omega_n=0)$ or,
equivalently, a jump of magnitude $C/T$ in
$\chi_{\rm loc}(T,\omega_n)$ as $\omega_n$ goes to zero.
(Its spin self-energy at zero frequency and zero temperature
tracks $I$.)
What this
implies is that the ``paramagnetic'' solution is not the physical
one; instead, the physical ground state corresponds to the magnetic
solution.
Nonetheless, the study of this unphysical paramagnetic solution is
helpful to the determination of the zero-temperature transition.
It is seen in Fig.~\ref{phase-diagram} that $C$ extrapolates to
zero at the same value of $I_c$ ($I \rightarrow I_c^+$) where
$E_{\rm loc}^\star$ goes to zero ($I \rightarrow I_c^-$). This
provides an additional consistency check for the critical
interaction where the paramagnetic phase terminates.

To summarize Fig.~\ref{phase-diagram}, within numerical accuracy,
all relevant scales vanish simultaneously at $I_c$, and the quantum
transition at zero temperature is continuous.

\subsection{Avoiding double-counting of the RKKY interaction}

To discuss the double-counting issue in some detail, we revisit the
procedure by which antiferromagnetism is treated in the standard
DMFT.~\cite{Georgesetal96} Here, the dynamical equations are
constructed entirely in terms of local single-particle quantities;
two-particle responses are derived once the dynamical equations have
been solved. On the paramagnetic side, the two-particle
susceptibility satisfies the following Bethe-Salpeter equation (in
matrix form):
\begin{eqnarray}
\chi^{-1} ({\bf q}, \omega) = \chi_{\rm p-h}^{-1} ({\bf q}, \omega)
- I_{\rm ir}({\bf q},\omega) . \label{chi-bethe-salpeter}
\end{eqnarray}
Here $\chi_{\rm p-h} ({\bf q}, \omega)$ is the particle-hole
susceptibility bubble of the full single-particle propagators
$G({\bf k}, \epsilon)$. The triplet particle-hole irreducible vertex
has the following form (again, in matrix form),~\cite{Zlatic90,Jarrell95}
\begin{eqnarray}
I_{\rm ir}({\bf q},\omega) = \chi_{\rm p-h,loc}^{-1} (\omega) -
\chi_{\rm loc}^{-1} (\omega) , \label{I-ir-dmft}
\end{eqnarray}
where $\chi_{\rm p-h,loc} (\omega)$ is the particle-hole
susceptibility bubble of the full local single-particle propagators
$G_{\rm loc}(\epsilon)$. Combining this with
Eq.~(\ref{chi-bethe-salpeter}) implies that, on the paramagnetic
side,
\begin{eqnarray}
\chi^{-1} ({\bf q}, \omega) =
\Delta I_{\bf q} + \chi_{\rm loc}^{-1} (\omega) .
\label{chi-pm-dmft}
\end{eqnarray}
where
\begin{eqnarray}
\Delta I_{\bf q} \equiv \chi_{\rm p-h}^{-1} ({\bf q}, \omega)
- \chi_{\rm p-h,loc}^{-1} (\omega)
\label{delta-I-q}
\end{eqnarray}
For our Kondo lattice model, $\Delta I_{\bf q}$ has the meaning of
the generated RKKY interaction [after inverting the matrix in the
$(f,c)$ space]. How can $\Delta I_{\bf q}$ appear in the
(particle-hole) spin response and not contribute in the dynamical
equations for the single-particle quantities? The answer lies in the
way the Brillouin zone is divided in DMFT into  ``special" ${\bf
q}$'s and ``generic" ${\bf q}$'s.~\cite{Muller-Hartmann} $\Delta
I_{\bf q}$ of Eq.~(\ref{delta-I-q}) is non-zero only at ``special''
${\bf q}$'s. The dynamical equations are constructed in terms of
quantities that are local, {\it i.e.} summed over ${\bf q}$: the
special ${\bf q}$'s, having measure zero, are not important for this
summation; only ``generic'' ${\bf q}$'s have the phase space to
contribute to the local quantities. To be more specific, consider
the hypercubic lattice. $\Delta I_{\bf q}$ depends on ${\bf q}$ only
through the combination $X({\bf q}) = (1/d) \sum_{\alpha=1}^{d} \cos
(q_{\alpha})$.~\cite{Muller-Hartmann} The dispersion $X({\bf q})$ is
$ O(1) $ (in the $d \rightarrow \infty$ limit) only for ``special''
${\bf q}$'s, ${\it e.g.}$ along the diagonals of the Brillouin zone,
$q_1=q_2=...=q_d$. On the other hand,  for ``generic'' ${\bf q}$,
$X({\bf q})$ vanishes [being formally of order $O(1/\sqrt{d})$, as
can be seen from the central-limit theorem]. When $X({\bf q})$
vanishes, $\chi_{\rm p-h} ({\bf q}, \omega) =\chi_{\rm p-h,loc}
(\omega)$, so $\Delta I_{\bf q}$ vanishes! The antiferromagnetic
wavevector {\bf Q} (Q$_{\alpha}=\pi$ for all $\alpha$), belongs to
the set of special ${\bf q}$'s, so $\Delta I_{\bf Q} \ne 0$. And the
antiferromagnetic instability, from the paramagnetic side, is
signaled by $\chi^{-1}({\bf Q},\omega)= \Delta I_{\bf Q} + \chi_{\rm
loc}^{-1} (\omega)=0$, at $\omega=0$. On the antiferromagnetic side,
the nonzero value for the corresponding $\Delta I_{{\bf Q},{\rm or}}$, is
implemented through the introduction of the doubling of the
conduction electron unit cell and different single-particle
propagators at the two sub-lattices. Formally, this doubling of the
conduction electron unit cell can be described in terms of an
effective susceptibility, ${\chi}_{\rm or} ({\bf Q}, \omega)$,
\begin{eqnarray}
{\chi_{\rm or}}^{-1} ({\bf Q}, \omega) =
\Delta I_{{\bf Q},{\rm or}} + \chi_{\rm loc}^{-1} (\omega) .
\label{chi-or-dmft}
\end{eqnarray}
Here, again, $\Delta I_{{\bf Q},{\rm or}}= \chi_{\rm p-h}^{-1} ({\bf
Q}, \omega) - \chi_{\rm p-h,loc}^{-1} (\omega)$. The instability of
the ordered state is signaled by ${\chi_{\rm or}}^{-1} ({\bf Q},
\omega) = \Delta I_{{\bf Q},{\rm or}} + \chi_{\rm loc}^{-1} (\omega)
=0$, also at $\omega=0$.
Because the effective RKKY interaction incorporated on the ordered
side, $\Delta I_{{\bf Q},{\rm or}}$, is the same as its counterpart
on the paramagnetic side, $\Delta I_{{\bf Q}}$,
the magnetic transition is in general of
second order. There is a major limitation to this approach. The RKKY
interaction, being zero at generic wavevectors, does not have enough
phase space to dynamically interplay with the Kondo interaction.
So the self-consistent dynamical equations of DMFT does not incorporate
$\Delta I_{\bf q}$ at all, and the Kondo screening is always present
including at the magnetic QCP. The quantum critical behavior falls
in the SDW category, the same as in any static mean-field description
of Kondo lattices.

The EDMFT is introduced precisely to allow this dynamical interplay.
Here, an intersite interaction, as given in $H_f$ of Eq.~(\ref{H-f}),
is elevated to the Hamiltonian level. The
Bethe-Salpeter equation~(\ref{chi-bethe-salpeter}) still applies.
However, the particle-hole irreducible vertex
becomes,~\cite{Smith2}
\begin{eqnarray}
I_{\rm ir}({\bf q},\omega) = \chi_{\rm p-h,loc}^{-1} (\omega) -
\chi_{\rm loc}^{-1} (\omega) -\chi_0^{-1}(\omega) - I_{\bf q}\;,
\label{I-ir-edmft}
\end{eqnarray}
where $I_{\bf q}$ is the Fourier transform of the intersite interactions,
$I_{ij}$, already included at the Hamiltonian level.
We have, on the paramagnetic side,
\begin{eqnarray}
\chi^{-1} ({\bf q}, \omega) = \Delta I_{\bf q} + I_{\bf q} +
M(\omega)\;, \label{chi-pm-edmft}
\end{eqnarray}
where $M(\omega) = \chi_{\rm loc}^{-1} (\omega) +
\chi_0^{-1}(\omega)$ is the spin self-energy. Likewise, we can write
the effective susceptibility that comes into the stability analysis
of the ordered phase as
\begin{eqnarray}
{\chi_{\rm or}}^{-1} ({\bf q}, \omega) = \Delta I_{{\bf q},{\rm or}}
+ I_{\bf q} + M(\omega)\;. \label{chi-or-edmft}
\end{eqnarray}

It was shown in Ref.~\onlinecite{Smith2} that the EDMFT can be
rigorously formulated only when all ${\bf q}$ are considered to
be generic. [Otherwise, $I_{\bf q}$ is formally of order
$O(d)$ at the special ${\bf q}$'s, and no paramagnetic phase could
exist.] This implies that $\Delta I_{\bf q} =0$ for all ${\bf
q}$. From the Kondo lattice point of view, this is equivalent to
saying that we will only use $I_{\bf q}$ to represent the RKKY
interaction and will not incorporate additional, generated RKKY
interactions from the fermion bubbles
(illustrated in Fig.~7 of
Ref.~\onlinecite{Smith2}).

In order to be consistent, one would also need to demand that
$\Delta I_{{\bf q},{\rm or}}=0$ on the ordered side. Otherwise, we
would be counting {\it additional} contributions to the RKKY
interaction on the ordered side that were absent on the paramagnetic
side. This requirement ($\Delta I_{{\bf q},{\rm or}}=0$) was
achieved in Ref.~\onlinecite{ZhuGrempelSi03} by using a featureless
conduction-electron band. 
The latter ensures that
all wavevectors are generic in the sense defined earlier.
Within this procedure, 
the magnetic ordering is entirely driven by $I_{\bf q}$,
and 
the instability
criteria from the paramagnetic and ordered sides coincide.
Therefore, the quantum transition is naturally of second order, as
was indeed seen numerically in Ref.~\onlinecite{ZhuGrempelSi03}; see
Fig.~\ref{phase-diagram} above.

The procedure used in Refs.~\onlinecite{SunKotliar03}
and~\onlinecite{SunKotliar05} amounts, in our language, to keeping
$\Delta I_{{\bf Q}}=0$ while $\Delta I_{{\bf Q},{\rm or}} \ne 0$.
On the paramagnetic side,
all wavevectors, including the 
ordering wavevector ${\bf Q}$,
are taken as generic, and
$\Delta I_{{\bf Q}}=0$, as in all EDMFT schemes.
On the ordered side,
${\bf Q}$ is considered as one of 
the special wavevectors in the sense defined earlier and,
through the conduction electron unit-cell doubling,
$\Delta I_{{\bf Q},{\rm or}} \ne 0$ 
(as in DMFT).
The ordered side then has an added energy gain, and the magnetic
quantum transition is of first order. 
(That an EDMFT approach to Kondo lattices which incorporates 
a DMFT-like fermion bubble on the ordered side alone yields a first order
transition at zero temperature was in fact already recognized
in \cite{SunKotliar05}.)
The procedure would
actually
lead to a first-order magnetic transition at zero-temperature in any
itinerant model, including  any $T=0$ SDW transition without any
Kondo physics.

We close by noting that the different EDMFT schemes that we have
discussed can be equivalently seen as different local approximations
to a Baym-Kadanoff-type functional.

\section{Experiments and Other Theoretical Approaches}

An important manifestation of the destruction of the
Kondo screening is that $f$-electrons participate in the Fermi
volume on the paramagnetic side but fails to do so on the
antiferromagnetic side. There is a sudden
reconstruction of the Fermi surface across the magnetic QCP. Fairly
direct electronic evidence for this effect has appeared in the
recent Hall-effect measurement in YbRh$_{\rm 2}$Si$_{\rm
2}$.~\cite{Paschen04} The Hall coefficient shows a rapid crossover as a
function of the control parameter --- magnetic field in this
case. The crossover sharpens as temperature is lowered,
extrapolating to a jump in the zero-temperature limit. The jump
occurs at the extrapolated location of the magnetic phase boundary
at zero temperature. Related
features have also been observed in YbAgGe.~\cite{Canfield05}

A more direct probe of the Fermi surface comes from the de Haas-van
Alphen effect. Recent dHvA measurement~\cite{Onuki} in CeRhIn$_{\rm
5}$ provides tantalizing evidence for a large reconstruction of the
Fermi surface, with a divergent effective mass, at a QCP.
Specific-heat measurement under magnetic field~\cite{Thompson05}
points towards the possibility that CeRhIn$_{\rm 5}$ undergoes a
second-order magnetic quantum transition at the magnetic field of
the strength used in the dHvA experiment. If the existence of 
the magnetic QCP is indeed established, CeRhIn$_{\rm 5}$ will 
provide more insights into quantum criticality than
CeRh$_{\rm 2}$Si$_{\rm 2}$. In the latter
system, a large Fermi-surface reconstruction has also been
seen in the dHvA measurements,~\cite{Onuki01} 
but the zero-temperature transition is likely to be first order
with a large jump in the magnetic order parameter
across the transition.

The fractional exponent and $\omega/T$ scaling in the magnetic
dynamics have been seen, since early on, in the antiferromagnetic
QCP of Au-doped CeCu$_{\rm 6}$~\cite{Schroeder00,Stockert98} (whose
magnetic fluctuations have a reduced dimensionality) and in some
frustrated 
compounds.~\cite{Aronson95,Wilson05} (On the other hand,
the SDW behavior is observed in the magnetic dynamics of
Ce(Ru$_{1-x}$Rh$_x$)$_2$Si$_2$,~\cite{Kadowaki05} which has quasi-3D
magnetic fluctuations.) Related non-trivial scaling exponents --
that are relatively easy to connect with theory -- have come from
the Gr\"{u}neisen ratio.~\cite{Kuchler03}

Theoretically, there have also been efforts to study the Kondo
lattice systems using certain mixed-boson-fermion representations
for the local-moment spin operators.~\cite{ColemanPepin03,Pepin05,Coleman05}
Such auxiliary-particle representations set up the basis for a picture
with spin-charge separation. However, it has been hard to use this
formalism to properly capture the Kondo-screened Fermi liquid
phase,~\cite{ParcolletGeorges97} making it  difficult to study
its destruction as well.

It may also be possible to describe the destruction of Kondo
screening in terms of the static mean field theories based on slave
boson and an RVB order parameter, supplemented by gauge-field
fluctuations. The corresponding phase diagram has recently been
studied in some detail.~\cite{Senthil04} The magnetic transition and
destruction of Kondo screening are found to occur at different
places in the zero-temperature phase diagram,~\cite{Senthil04} so
the magnetic quantum transition is still of the SDW type. We believe
that this is a reflection of the static nature of the mean field
theory.

Finally, it is instructive to put in the present context the QCP
proposed for the transition from an antiferromagnet to a
valence-bond solid in frustrated 
quantum
magnets.~\cite{Senthil-Science04}
Dubbed a ``deconfined'' QCP, it has certain properties that may be
qualitatively compared with the local quantum criticality: the QCP
-- containing exotic excitations -- is surrounded by two
conventional phases, and the corresponding energy scales of both
vanish as the QCP is approached. Hence, it would be enlightening to
explore the concrete connections (if any) of this approach with the
physics of the destruction of Kondo screening. For this purpose, it
would be necessary to either construct microscopic spin models for
the deconfined QCP or reformulate the Kondo screening beyond
microscopic approaches.

\section{Summary and Outlook}

To summarize, we have discussed some of the microscopic approaches underlying
the local quantum critical picture.
Beyond the initial studies based on an $\epsilon-$expansion
renormalization group method, the most extensive
investigations have been carried out in Kondo lattice
models with Ising anisotropy. The latter have allowed
the study of both the destruction of Kondo screening and
the concomitant fractional exponent and $\omega/T$
scaling in the magnetic dynamics. We have also discussed
the magnetic phase diagram and summarized the evidence
for the second order nature of the magnetic quantum phase
transition. The EDMFT studies of Kondo lattice models
with continuous spin symmetry [$SU(2)$ or $XY$] are mostly confined
to the $\epsilon$-expansion studies. Efforts to access the quantum
critical point in these systems, beyond the $\epsilon$-expansion,
are still underway.
A dynamical large-$N$ limit, for instance, has recently been shown
to be promising.~\cite{ZhuKirchner04}

The microscopic approaches described here have shown that critical
modes beyond the order parameter fluctuations
exist, and the modes are associated with the destruction
of Kondo screening. These insights have a number of phenomenological
consequences -- not only in magnetic dynamics but also in the
Fermi surface properties and in thermodynamics -- which have been
supported by the existing and emerging experiments. The insights
will also help the search for the field theory that describes
quantum critical heavy fermions. Finally, they may very well be
broadly relevant to the exotic quantum critical behavior in other
strongly correlated systems such as doped Mott insulators.


We would particularly like to thank S. Burdin, M. Grilli,
K. Ingersent, S. Kirchner, E. Pivovarov, and L. Zhu
for collaborations on related work, and
A. Georges, G. Kotliar, and T. Senthil for discussions.
One of us (QS) would like to thank C. P{\'e}pin for her organization
of an informal workshop at Saclay (Dec. 2004), where some
of these issues were discussed. We are especially grateful
to G. Kotliar and P. Sun for pressing us to clarify
the issue of the magnetic phase diagram. This work has been
supported in part by NSF Grant No.\ DMR-0424125 and the Robert
A. Welch foundation (QS), and the US DOE (JXZ).

\end{document}